\documentclass{emulateapj}
\usepackage{graphicx}

\def\kms    {\ifmmode{{\rm \ts km\ts s}^{-1}}\else{\ts km\ts s$^{-1}$}\fi}
\def\msol   {\ifmmode{{\rm M}_{\odot}}\else{M$_{\odot}$}\fi}
\def\lsun   {\ifmmode{{\rm L}_{\odot}}\else{L$_{\odot}$}\fi}
\def\ts     {\thinspace} 
\def\ci   {\ifmmode{{\rm C}{\rm \small I}}\else{C\ts {\scriptsize I}}\fi}
\def\cone {\ifmmode{{\rm C}{\rm \small I}(1-0)}\else{C\ts {\scriptsize I}(1--0)}\fi}
\def\ctwo {\ifmmode{{\rm C}{\rm \small I}(2-1)}\else{C\ts {\scriptsize I}(2--1)}\fi}
\def\cii  {\ifmmode{{\rm [C}{\rm \small II}]}\else{[C\ts {\scriptsize II}]}\fi}
\def\aco  {\ifmmode{^{12}{\rm CO}(J=1\to0)}\else{$^{12}{\rm CO}(J=1\to0)$}\fi}
\def\bco  {\ifmmode{^{12}{\rm CO}(J=2\to1)}\else{$^{12}{\rm CO}(J=2\to1)$}\fi}
\def\m    {\ifmmode{\mu {\rm m}}\else{$\mu$m}\fi}
\def\cco  {\ifmmode{^{13}{\rm CO}(J=1\to0)}\else{$^{13}{\rm CO}(J=1\to0)$}\fi}
\def\dco  {\ifmmode{^{13}{\rm CO}(J=2\to1)}\else{$^{13}{\rm CO}(J=2\to1)$}\fi}
\def\eco  {\ifmmode{^{12}{\rm CO}(J=3-2)}\else{$^{12}{\rm CO}(J=3-2)$}\fi}
\def\hi   {\ifmmode{{\rm H}{\rm \small I}}\else{H\ts {\scriptsize I}}\fi}
\def\hii  {\ifmmode{{\rm H}{\rm \small II}}\else{H\ts {\scriptsize II}}\fi}
\def\ha   {\ifmmode{{\rm H}{\alpha}}\else{H${\alpha}$}\fi}
\def\hh     {\ifmmode{{\rm H}_2}\else{H$_2$}\fi}
\def\nhh     {\ifmmode{N({\rm H}_2)}\else{$N$(H$_2$)}\fi}
\def\tex {\ifmmode{{T}_{\rm ex}}\else{$T_{\rm ex}$}\fi}
\def\tmb {\ifmmode{{T}_{\rm mb}}\else{$T_{\rm mb}$}\fi}
\def\tkin {\ifmmode{{T}_{\rm kin}}\else{$T_{\rm kin}$}\fi}
\def\microns {\ifmmode{\mu{\rm m}}\else{$\mu$m}\fi}
\def\nhh   {\ifmmode{n({\rm H}_2)}\else{$n$(H$_2$)}\fi}

\shorttitle{Evidence for low extinction in actively star forming galaxies at z$>$6.5.}

\shortauthors{Walter et al.}

\journalinfo{Accepted for publication in the Astrophyscial Journal}
\begin{document}

\title{Evidence for Low Extinction in Actively Star Forming Galaxies at z$>$6.5.}
   \author{
          F. Walter
          \altaffilmark{1},
          R. Decarli
          \altaffilmark{1},
          C. Carilli
          \altaffilmark{2},
          D. Riechers
          \altaffilmark{3},
          F. Bertoldi
          \altaffilmark{4},
          A. Wei{\ss}
          \altaffilmark{5},
          P. Cox
          \altaffilmark{6},
          R. Neri
          \altaffilmark{6},
          R. Maiolino
          \altaffilmark{7},
          M. Ouchi
          \altaffilmark{8},
          E. Egami
          \altaffilmark{9},
          K. Nakanishi
          \altaffilmark{10,11,12}
           }
\altaffiltext{1}{Max-Planck-Institut f\"ur Astronomie, K\"onigstuhl 17, D-69117 Heidelberg, Germany    [e-mail: {\em walter@mpia.de}]}
\altaffiltext{2}{National Radio Astronomy Observatory, 1003 Lopezville Road, Socorro, NM 87801, USA}
\altaffiltext{3}{California Institute of Technology, Pasadena, CA}
\altaffiltext{4}{Argelander-Institut f\"ur Astronomie, Auf dem H\"ugel 71, D--53121 Bonn, Germany}
\altaffiltext{5}{Max-Planck-Insitut f\"ur Radioastronomie, Auf dem H\"ugel 69, D-53121 Bonn, Germany}
\altaffiltext{6}{Institut de Radio Astronomie Millim\'etrique (IRAM), St. Martin d'H\`eres, France}
\altaffiltext{7}{Institute of Astronomy, Cambridge, UK}
\altaffiltext{8}{University of Tokio, Tokio, Japan}
\altaffiltext{9}{University of Arizona, Tucson, AZ, USA}
\altaffiltext{10}{ALMA Project Office, National Astronomical Observatory, 2-21-1 Osawa, Mitaka, Tokyo 181-8588, Japan}
\altaffiltext{11}{Joint ALMA Office, Alonso de Cordova 3107, Vitacura, Santiago 763 0355, Chile}
\altaffiltext{12}{The Graduate University for Advanced Studies (Sokendai), 2-21-1 Osawa, Mitaka, Tokyo 181-8588, Japan}
\begin{abstract} 

We present a search for the \cii\ 158$\mu$m fine structure line (a
main cooling line of the interstellar medium) and the underlying
far--infrared (FIR) continuum in three high--redshift (6.6$<$z$<$8.2)
star--forming galaxies using the IRAM Plateau de Bure
interferometer. We targeted two Lyman--$\alpha$--selected galaxies
(Lyman--Alpha--Emitters, LAEs) with moderate UV--based star formation
rates (SFR$\sim$20\,M$_\odot$\,yr$^{-1}$; {\it Himiko} at z=6.6 and
{\it IOK--1} at z=7.0) and a Gamma Ray Burst (GRB) host galaxy ({\it
GRB\,090423} at z$\sim$8.2). Based on our 3$\sigma$ rest--frame FIR
continuum limits, previous (rest--frame) UV continuum measurements and
spectral energy distribution (SED) fitting, we rule out SED shapes
similar to highly obscured galaxies (e.g. Arp\,220, M\,82) and less
extreme dust--rich nearby spiral galaxies (e.g. M\,51) for the
LAEs. Conservatively assuming a SED shape typical of local spiral
galaxies we derive upper limits for the FIR--based star formation
rates (SFRs) of $\sim$70\,M$_\odot$\,yr$^{-1}$,
$\sim$50\,M$_\odot$\,yr$^{-1}$ and $\sim$40\,M$_\odot$\,yr$^{-1}$ for
{\it Himiko}, {\it IOK--1} and {\it GRB\,090423}, respectively. For
the LAEs these limits are only a factor $\sim$3 higher than the
published UV--based SFRs (uncorrected for extinction). This indicates
that the dust obscuration in the z$>$6 LAEs studied here is lower by a
factor of a few than what has recently been found in some LAEs at
lower redshift (2$<$z$<$3.5) with similar UV--based SFRs. A low
obscuration in our z$>$6 LAE sample is consistent with recent
rest--frame UV studies of z$\sim$7 Lyman--Break--Galaxies (LBGs).

\end{abstract}


\section{Introduction}

Characterizing the physical properties of the earliest galaxies in the
universe is a key goal in observational astrophysics. Of particular
interest is the redshift range z$>$6 (corresponding to the first Gyr
of the universe), in which the universe underwent a phase transition
from a mostly neutral to a mostly ionized universe (the `Epoch of
Reionization', see, e.g., reviews by Fan, Carilli \& Keaton 2006,
Robertson et al.\ 2010). Characterizing the physical properties of
galaxy populations in this early epoch, and their respective
contribution to reionization, are important drivers in current studies
of the high redshift universe. Recent deep dropout--studies in the
Hubble Ultra Deep Field using the Hubble Space Telescope have revealed
a population of z$>$6.5 galaxy candidates, out to possibly z$\sim$10
(e.g., Bouwens et al.\ 2011).  Given their faintness in optical and
near--infrared (NIR) bands, spectroscopic confirmation of these
sources has been challenging, leading to only a few confirmed galaxies
at z$\gtrsim$7.0 (e.g., Fontana et al.\ 2010, Vanzella et al.\ 2011,
Kashikawa et al.\ 2011, Ono et al.\ 2012, Pentericci et al.\ 2011,
Schenker et al.\ 2011).

Galaxies with the highest Ly--$\alpha$--based SFRs at these redshifts
(z$>$6.5) are selected through wide--field, narrow--band surveys, the
so-called `Lyman--$\alpha$ Emitters' (hereafter LAEs). A high fraction
of narrow--band selected candidates is  spectroscopically
confirmed to indeed lie at z$>$6.5 (e.g., Taniguchi et al.\ 2005, Iye
et al.\ 2006, Ouchi et al.\ 2009, Ouchi et al.\ 2010, Kashikawa et
al.\ 2011, Jiang et al.\ 2011).  Their Ly--$\alpha$ fluxes imply SFRs
of typically $\sim$10\,M$_{\odot}$\,yr$^{-1}$, consistent with their
(rest--frame) UV luminosities. A recent study of low--z (z$\sim$0.3)
LAEs by Oteo et al. (2012a) with lower UV--based SFRs find that
obscured star formation contributes to more than 50\% of the total
SFR, as traced by both UV and FIR emission. In a higher redshift
(2.0$<$z$<$3.5) sample of LAEs, Oteo et al.\ (2012b) report that some
FIR--based SFRs (for objects with UV--based SFRs comparable to the
sources studied here) contribute more than 90\% of the total SFRs
(their Tab.~1).  This implies that the presence of detectable dust and
Ly--$\alpha$ emission is not mutually exclusive in these objects (Oteo
et al.\ 2012b).

An alternative way to pinpoint the locations of active star formation
in the very early universe are Gamma Ray Bursts (GRBs) that are thought to
be a common phenomenon at redshifts z$>$6.5 (e.g., review by Woosley \&
Bloom 2006, Bromm \& Loeb 2006, Salvaterra \& Chincarini
2007). Observations indicate that the GRB rate may evolve even more
rapidly than the star formation history of the universe up to z=4
(Kistler et al.\ 2008, Y\"uksel et al.\ 2008). Since the progenitors
of long GRBs are thought to be short--lived massive stars of one of
the Universe's earliest stellar generation, their detection at extreme
redshifts offers the possibility to put approximate constraints on the
star formation rates in their host galaxies.

Although z$>$6.5 galaxies are key to understanding early galaxy
formation and reionization, it is a major challenge to study them at
their extreme redshifts. Even with long (tens of hours) exposures on
8--10\,m class telescopes, optical/NIR observations can only detect
the Ly--$\alpha$ line and the underlying UV stellar continuum.
Indeed, essentially all of our current knowledge on these key
reionization sources is based on the highly resonant Ly--$\alpha$ line
and the faint underlying continuum, which is potentially affected by
dust attenuation, and generally difficult to interpret due to the
complexity of the Ly--$\alpha$ radiative transfer.

Observations of the (sub--)millimeter continuum and spectral lines in
principle provide the means by which to constrain key physical
parameters of the highest redshift sources (e.g., size, mass,
distribution of the star--forming interstellar medium (ISM) and star
formation, as well as dynamical masses).  Traditionally, CO emission
lines are used to characterize the molecular reservoirs in high
redshift galaxies (e.g. reviews by Solomon \& Vanden Bout 2005,
Walter, Carilli \& Daddi 2011). However, the expected CO line
strengths even in the brightest z$>$6.5 (non--AGN) galaxies are
$\sim$10s of $\mu$Jy --- such faint lines are not accessible given
current facilities and will even be a challenge with ALMA. On the
other hand the $^2$P$_{3/2}$ $\to$ $^2$P$_{1/2}$ fine--structure line
of \cii\ (a major cooling line of the ISM) is expected to be much
brighter than the CO lines: It has long been known (e.g.  Stacey et
al.\ 1991, Malhotra et al.\ 1997), that \cii\ is tracing star
formation (photon dominated regions, PDRs) and that it can carry up to
1\% of the total far--infrared emission of a galaxy, in particular in
systems of low luminosity and metallicity (e.g., Israel et al.\ 1996,
Madden et al.\ 1997).  The high ratio of L$_{\rm [CII]}$/L$_{\rm FIR}$
(see Sec.~3) is the reason why it has been argued for more than a
decade that observation of the \cii\ line of pristine systems at the
highest redshifts will likely be the key to study the star--forming
ISM in the earliest star--forming systems (e.g., Stark 1997, Walter \&
Carilli 2008, Carilli et al.\ 2008). Indeed, the last few years have
seen a steep incfrease in the number of high-z (2$<$z$<$6.5) \cii\
detections, in most cases in systems that host an active galactic
nucleus (AGN) and/or very high SFRs ($>>$100\,M$_\odot$\,yr$^{-1}$,
e.g.  Maiolino et al.\ 2005, Maiolino et al.\ 2009, Walter et al.\
2009, Stacey et al.\ 2010, Ivison et al.\ 2010, Wagg et al.\ 2010, Cox
et al.\ 2011, De Breuck et al.\ 2011, Valtchanov et al.\ 2011).

\cii\ is currently the most promising tracer of the star--forming ISM
(both PDRs and the cold neutral medium, CNM) in galaxies at the
highest redshifts. However, a calibration to derive meaningful star
formation rates from L$_{\rm [CII]}$ is still lacking. The \cii\ line
also traces the CNM and thus, is a key tracer for the overall
distribution of the ISM and its global dynamics (and thus dynamical
masses). In this paper we present the results of a search for \cii\
emission and the underlying FIR continuum in some of the
highest--redshift (non--AGN) galaxies with moderate UV--based SFR
($\sim$20\,M$_\odot$\,yr$^{-1}$). In Sect.~2 we describe the target
selection and observations. The data are shown in Sect.~3 and
implications are summarized in Sect.~4. Throughout this paper we use a
$\Lambda$--Cold Dark Matter cosmology with $H_{\rm 0} = 70$
\kms\,Mpc$^{-1}$, $\Omega_\Lambda=0.7$ and $\Omega_m=0.3$.

\begin{figure} \centering
\includegraphics[width=9.0cm,angle=0]{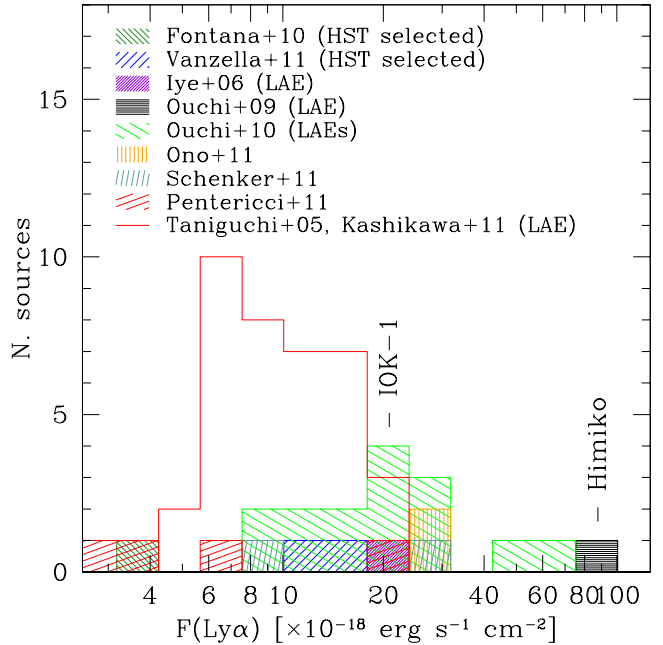} \caption{Distribution
of Ly--$\alpha$ fluxes (a proxy for star formation) in all published
spectroscopically confirmed z$>$6.5 LAEs and LBGs (Taniguchi et al.\ 2005, Iye et
al.\ 2006, Ouchi et al.\ 2009, Ouchi et al.\ 2010, Fontana 
et al.\ 2010, Vanzella et al.\
2011, Kashikawa et al.\ 2011, Ono et al.\ 2011, Pentericci et al.\
2011, Schenker et al.\ 2011). Most of these galaxies are too faint for
detection in \cii\ given current facilities. The targets selected in
this study ({\it Himiko} and {\it IOK--1}) are amongst the brightest
sources. {\it IOK--1} is preferred over other LAEs of similar
luminosity given its slightly higher redshift (z$\sim$7.0) and thus
better receiver performance at the redshifted \cii\ frequency.}  \end{figure}

\begin{figure*} \centering
\includegraphics[width=16.0cm,angle=0]{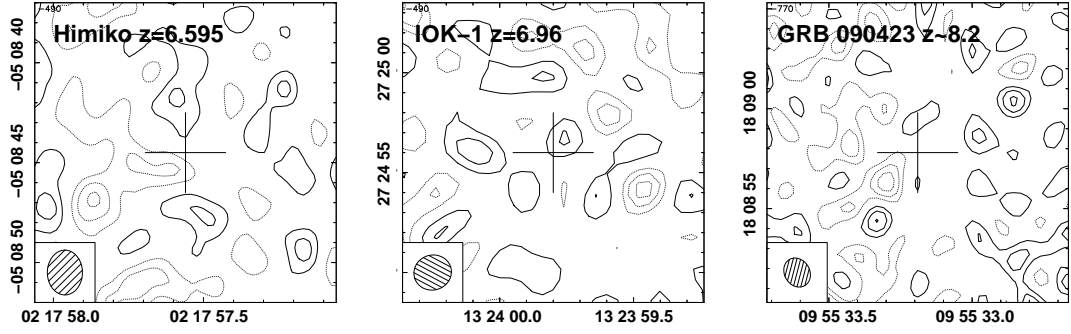}\vspace{-3mm} \caption{Rest--frame
158$\mu$m continuum observations of {\it Himiko} ({\it left}), {\it
IOK--1} ({\it middle}) and {\it GRB\,090423} ({\it right
panel}). Coordinates are given in the J2000.0 system and contours are shown in steps of 1\,$\sigma$, starting at
$\pm$1\,$\sigma$ (0.21\,mJy\,beam$^{-1}$, 0.15\,mJy\,beam$^{-1}$ and
0.089\,mJy\,beam$^{-1}$, respectively, see Tab.~1). No significant emission is detected from any of the sources.}  \end{figure*}

\section{Observations}

\subsection{Source Selection} 

\subsubsection{Lyman Alpha Emitters} 

Out of the many dozens of spectroscopically confirmed z$>$6.5 LAEs
(see Fig.~1 caption for references) we have selected two targets with
high SFRs based on their Ly--$\alpha$ and (rest--frame) UV continuum
luminosity. The first source is an exceptionally luminous LAE at
z=6.595 (Ouchi et al.\ 2009) dubbed {\it Himiko} by these authors.
{\it Himiko} is by far the brightest and most spatially extended ($\sim3"$) LAE
known at these redshifts, with a lower limit to the SFR derived from
the rest--frame UV emission (with no correction for dust attenuation)
of $\sim$25\,M$_{\odot}$\,yr$^{-1}$.  {\it Himiko} is even detected at
3.6\,$\mu$m using Spitzer, implying a significant underlying stellar
mass of $\sim$4$\times$10$^{10}$\,M$_\odot$ (Ouchi et al.\ 2009).

The second source is {\it IOK--1}, a LAE at z=6.96 (Iye et al.\
2006). This LAE is similar in both Ly--$\alpha$ and continuum
brightness to those in other samples at z$\sim$6.6 (see Fig.~1). The
(rest--frame) UV emission implies a SFR of
$\sim16$\,M$_\odot$\,yr$^{-1}$ and HST imaging of {\it IOK--1} gives a
source size of $<1"$ (Cai et al.\ 2011).

\subsubsection{{\it GRB\,090423}}

The third source is {\it GRB\,090423} (Krimm et al.\ 2009).  VLT
spectroscopy of {\it GRB\,090423} gave a redshift of
8.26$^{+0.07}_{-0.08}$, based on a continuum break due to
Gunn--Peterson absorption (Tanvir et al.\ 2009). A similar estimate
was reported by Salvaterra et al.\ (2009) using spectroscopy at the
TNG (z=8.1$^{+0.1}_{-0.3}$).  This makes {\it GRB\,090423} amongst the
highest redshift GRBs discovered. The afterglow has been detected in
the millimeter continuum by the IRAM Plateau de Bure interferometer
(PdBI) at 3\,mm wavelengths with a flux density of 0.3$\pm$0.1\,mJy
(Castro--Tirado et al.\ 2009, de Ugarte Postigo et al.\ 2012, obtained
on 2009, April 23 \& 24). An upper limit of the afterglow and the host
galaxy FIR flux density at 1.2\,mm wavelengths of 0.23$\pm$0.32mJy
using MAMBO has also been reported (Riechers et al.\ 2009, obtained on
2009, April 25).

\begin{deluxetable*}{lllccrrlrrr}
\tabletypesize{\scriptsize}
\tablecaption{Summary of Observations and Derived Properties}
\tablewidth{0pt}
\tablehead{
\colhead{source\tablenotemark{a}} & \colhead{RA} & \colhead{DEC}  & \colhead{z\tablenotemark{a}} & \colhead{$\nu$\tablenotemark{b}} & \colhead{$\sigma_{\rm cont}$\tablenotemark{c}} & \colhead{$\sigma_{\rm line}$\tablenotemark{d}}  & \colhead{L$_{\rm [CII]}$\tablenotemark{e}} & \colhead{L$_{\rm IR}^{\rm N6946}$} & \colhead{SFR$^{\rm dust}$\tablenotemark{f}} & \colhead{SFR$^{\rm UV}$\tablenotemark{g}}\\
\colhead{}    & \colhead{J2000.0}  & \colhead{J2000.0}&  \colhead{}   & \colhead{GHz} & \colhead{mJy\,b$^{-1}$}   &  \colhead{mJy\,b$^{-1}$} &  \colhead{10$^8$\,L$_{\odot}$}  &  \colhead{10$^{11}$\,L$_{\odot}$} & \colhead{M$_{\odot}$\,yr$^{-1}$} & \colhead{M$_{\odot}$\,yr$^{-1}$ }
}
\startdata
{\it Himiko}     & 02:17:57:56 & --05:08:44.5 & 6.595    & 250.361        & 0.21  & 0.70  & $<$4.43 & $<$3.98 &  $<$69 & $\sim$25  \\
{\it IOK--1}     & 13:23:59.80 & +27:24:56.0  & 6.96     & 238.881        & 0.15  & 0.65  & $<$4.42 & $<$3.02 &  $<$52 & $\sim$16  \\
{\it GRB\,090423}& 09:55:33.19 & +18:08:57.8  & $\sim$8.2& 203.38, 206.98 & 0.089 & 0.61  & $<$5.44 & $<$2.29 &  $<$39 & \nodata 
\enddata
\tablecomments{All luminosities upper limits are 3\,$\sigma$.}
\tablenotetext{a}{References: {\it Himiko}: Ouchi et al. 2009, {\it IOK--1}: Iye et al. 2006, {\it GRB\,090423}: Tanvir et al.\ 2009, Salvaterra et al.\ 2009.}
\tablenotetext{b}{Frequencies for {\it Himiko} and {\it IOK--1} are tuned $\sim$125\,MHz blueward of Ly--$\alpha$ redshift.}
\tablenotetext{c}{1$\sigma$ continuum sensitivity at 158$\mu$m rest wavelengths.}
\tablenotetext{d}{1$\sigma$ \cii\ line sensitivity over a channel width of 200\,km\,s$^{-1}$.}
\tablenotetext{e}{3$\sigma$ \cii\ luminosity limit over a channel width of 200\,km\,s$^{-1}$ assuming $L_{\rm line} = 1.04 \times 10^{-3} \, F_{\rm line} \, \nu_{\rm rest} (1+z)^{-1}\, D_{\rm L}^{2}$, where the line luminosity, $L_{\rm line}$, is measured in $L_\odot$; the velocity integrated flux, $F_{\rm line}$=$S_{\rm line}\, \Delta v$, in Jy\,\kms; the rest frequency, $\nu_{\rm rest} = \nu_{\rm obs} (1+z)$, in GHz; and the luminosity distance, $D_{\rm L}$, in Mpc.}
\tablenotetext{f}{3$\sigma$ limit based on L$_{\rm IR}^{\rm N6946}$.}
\tablenotetext{g}{UV--based SFR from Ouchi et al.\ (2009) and Cai et al.\ (2011).}
\end{deluxetable*}

\begin{figure}\includegraphics[width=9cm,angle=0]{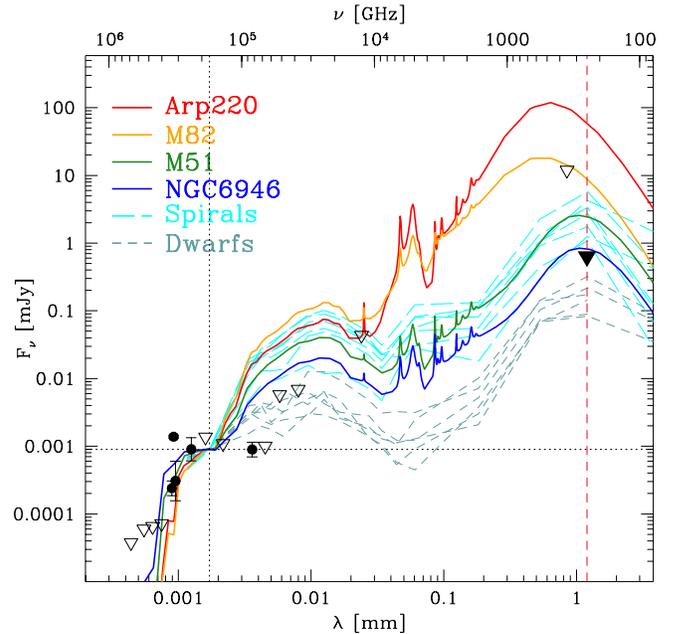}
\caption{SED of {\it Himiko} in the observer's frame (bottom:
wavelength, top: frequency). Detections in  the (rest--frame) UV and B--band 
are indicated as black points (Ouchi et al.\ 2009). 3--$\sigma$ upper
limits from Ouchi et al.\ (2009) are shown as open triangles. Our
3--$\sigma$ upper limit at 158$\mu$m (rest--frame, see vertical red line) 
is shown as a black
triangle. A number of galaxy SEDs are shown for comparison (see color
coding in the panel) normalized to the UV measurement longward of
Lyman--$\alpha$ (as indicated by the vertical and horizontal dashed lines). 
The SEDs of Arp\,220, M\,82, M\,51 and NGC\,6946 are
from Silva et al.\ (1998) and the cyan and grey curves show individual
measurements of nearby spiral and dwarf galaxies, respectively (Dale
et al.\ 2007). Our continuum limit rules out most spiral galaxy (and
dustier system) SED shapes but is compatible with the SED shapes
representative of lower dust content found in nearby dwarf galaxies.
} \end{figure}

\subsection{PdBI Observations}

{\it Himiko}, {\it IOK--1} and {\it GRB\,090423} have been observed
with the IRAM Plateau de Bure interferometer using the WideX wide
bandwidth correlator. The \cii\ line (rest frequency: 1900.54\,GHz,
157.74$\mu$m) is shifted to the 1\,mm band and we have tuned the
receivers 150\,km\,s$^{-1}$ ($\sim$125\,MHz) blueward of the redshift
derived from optical spectroscopy (see Tab.~1 for the exact tuning
frequencies). The total bandwidth of WideX is 3.6\,GHz, or
$\sim$4400\,km\,s$^{-1}$ at the observed frequencies. Thus any
conceivable shift between the Ly--$\alpha$ and the \cii\ lines are
covered by our observations. Observations were typically carried out
during good observing conditions, using standard calibrations. {\it
IOK--1} was observed during 3 runs (April--October 2010) for an
equivalent (5--element interferometer) on--source time of 9.1\,h ({\it
Himiko}: 7 days from August--December 2010, total of 11.5\,h, {\it
GRB\,090423}: 4 days in April 2010, total of 9.7\,h). The following
gain calibrators were used: {\it Himiko}: B0336--019, {\it IOK--1}:
B1308+326, {\it GRB\,090423}: B1040+244.

Table~1 summarizes the sensitivity reached in the observations; here
we list the 1$\sigma$ continuum sensitivity as well as the 1$\sigma$
line sensitivity over a 200\,km\,s$^{-1}$ channel (which we adopt as
realistic line widths of these sources, given their Ly--$\alpha$
widths). The observations resulted in the following beamsizes: {\it
Himiko}: $2.27"\times 1.73"$, PA=172$^\circ$ (C and D configurations),
{\it IOK--1}: $1.88" \times 1.75"$, PA=109$^\circ$ (D configuration)
and {\it GRB\,090423}: $1.51" \times 1.22"$, PA=30.29$^\circ$ (C
configuration). {\it Himiko} has a spatial extent that is similar to
our beamsize ($\sim3"$, Ouchi et al.\ 2009), so we consider the risk
of outresolving the source as small.\\

\section{Results and Discussion}

\subsection{Continuum Emission} 

None of our sources is detected in the rest--frame FIR continuum
(Fig.~2). In Fig.~3 we overplot our 3--$\sigma$ continuum limit for
{\it Himiko}, for which the best UV/optical data exist to date (Ouchi
et al.\ 2009), on top of galaxy spectral energy distribution (SED)
templates. The templates include Arp\,220, M\,82, M\,51 and NGC\,6946
(Silva et al.\ 1998), as well as spiral (cyan) and dwarf (blue) galaxy
SEDs from the observations presented in Dale et al.\ (2007). All SEDs
are shifted to {\it Himiko}'s redshift z=6.595 and are normalized to
the (rest--frame) UV continuum, shortly longward of the Ly--$\alpha$
wavelength. Based on our  continuum upper limit we can rule out a
number of galaxy SED shapes. In particular, our measurement is
(perhaps not surprisingly) incompatible with very obscured dusty
starburst templates, such as Arp\,220 and M\,82, but also dust--rich
nearby spiral galaxies such as M\,51.  For {\it IOK--1} (not shown)
the situation is very similar: the source has a similar redshift,
millimeter continuum limit, and UV--based SFR (see Tab.~1).

Based on Fig.~3 we would have detected the dust continuum at
$>$3\,$\sigma$ if {\it Himiko's} SED resembled M\,51 or NGC\,6946, but
note that such SEDs (and those of more metal--poor dwarfs) are also
already excluded from the IRAC 3.6\,$\mu$m measurement which
corresponds to a rest wavelength of 475\,nm. This flux measurement,
as discussed in detail in Ouchi et al.\ (2009), is dependent on the
presumably young stellar component that is in place at z=6.6. As the
relation of such a pristine stellar population to the dust emission,
and the relative spatial distribution between the two, is unknown, the
FIR limit is still relevant. In any case our observations point
towards a dust--poor environment, e.g. as seen in the case of local
dwarf galaxies (see dwarf galaxy SEDs in Fig.~3).

Fig.~3 shows that our continuum measurement is close in wavelength to
the peak of any likely galaxy SED (spiral/dwarfs). In the following we
derive a conservative upper limit for L$_{\rm IR}$ (integrated from
8--1000$\mu$m) using the galaxy template for NGC\,6946, scaled to our
3--$\sigma$ upper limit (we consider this limit conservative as using
a dwarf galaxy template would decrease this upper limit). From this we
derive 3$\sigma$ upper limits of the (F)IR--based SFR (using
SFR(M$_\odot$\,yr$^{-1}$)=1.72$\times$10$^{-10}$ L$_{\rm IR}$
(L$_\odot$), Kennicutt 1998) of $\sim$70\,M$_\odot$\,yr$^{-1}$ ({\it
Himiko}) and $\sim$50\,M$_\odot$\,yr$^{-1}$ ({\it IOK--1}, see
Tab.~1). We note that our upper limit of the FIR--based SFR for {\it
Himiko} and {\it IOK--1} is only a factor $\sim$3 higher than the
UV--based SFR (not corrected for exctinction).

In the case of the {\it GRB\,090423} host galaxy we derive an upper
limit of its FIR--based SFR$<$39\,M$_\odot$\,yr$^{-1}$ using the same
SED template as above.  Our SFR limits for {\it GRB\,090423} is the
deepest FIR--based SFR measurement of a GRB to date (see e.g. Berger
et al.\ 2003, de Ugarte Postigo et al.\ 2011, Hatsukade et al.\
2011). Very recent deep HST observations did not reveal the host
galaxy of {\it GRB\,090423} either down to a limiting UV--based SFR of
0.4\,M$_\odot$\,yr$^{-1}$ (Tanvir et al. 2012).

\subsection{\cii\ Line}

As in the case of the continuum, no significant \cii\ emission is
detected in our observations (see limits in Tab.~1). In Fig.~4 we show
the channel maps around the expected Lyman--$\alpha$ redshift for {\it
Himiko}.  We note a tentative (3\,$\sigma$) signal 1.5$"$
(0.7$\times$FWHM$_{\rm maj}$) south of the phase center, close to the
central tuning frequency, but more sensitive observations are needed
to further investigate this finding.

\begin{figure} \centering
\includegraphics[width=9cm,angle=0]{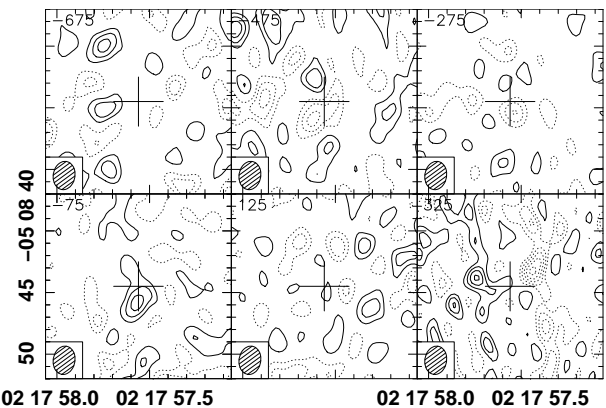} \caption{{\it Himiko}
\cii\ velocity channel maps (width: 200\,km\,s$^{-1}$) around the expected
redshift. Contours are shown in
steps of 1\,$\sigma$ (0.70\,mJy\,beam$^{-1}$) starting at
$\pm$1\,$\sigma$.  Central velocities are shown in the top left corner
of each panel. Note that due to our tuning the systemic redshift
corresponds to a velocity of --150\,km\,s$^{-1}$ in this
representation (see Sec.~2.2). A tentative 3$\sigma$ signal is seen
$\sim$1.5$"$ away from the phase center in the bottom left panel.  }  \end{figure}

\begin{figure} \centering
\includegraphics[width=9.0cm,angle=0]{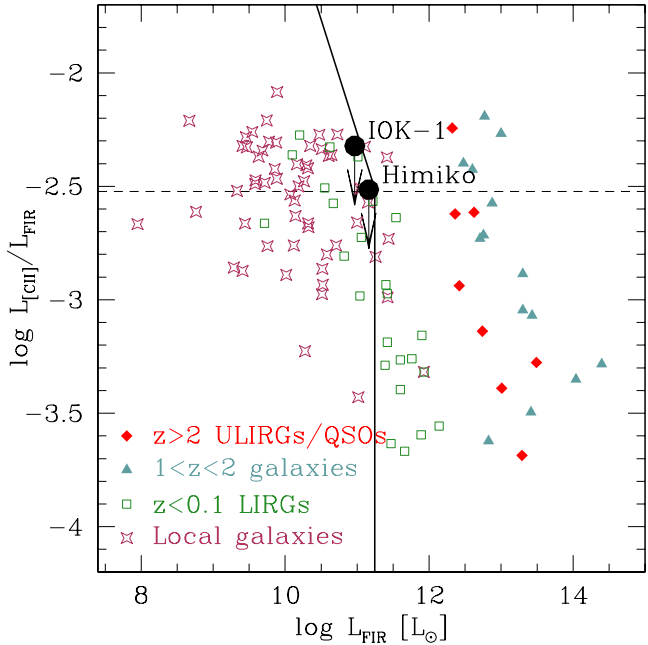} \caption{L$_{\rm
[CII]}$/L$_{\rm FIR}$ as a function of L$_{\rm FIR}$ for published
\cii\ detections (Malhotra et al.\ 2001, Maiolino et al.\ 2009, Stacey
et al.\ 2010, Ivison et al.\ 2010, Wagg et al.\ 2010, Cox et al.\
2011, De Breuck et al.\ 2011). As neither FIR nor [CII] emission
is detected in our targets, the shaded area indicates the possible
parameter for {\it Himiko} and {\it IOK--1} based on the upper
limits. Assuming that the FIR--based (extincted) SFR equals the
UV--based (unextincted) SFR (i.e. a FIR luminosity that is three times
lower than our upper limit) results in the two upper limit datapoints
shown for {\it Himiko} and {\it IOK--1}. The horizontal dashed line 
indicates the average value of L$_{\rm
[CII]}$/L$_{\rm FIR}$=3$\times$10$^{-3}$ for local galaxies.}
  \end{figure}

In Fig.~5 we plot the ratio L$_\cii$/L$_{\rm FIR}$ as a function of
L$_{\rm FIR}$ for objects taken from the literature (see references in
the caption). Given our limits on the FIR and [CII] luminosities we
can now constrain the location of our targets in this plot. The likely
region in parameter space is shown as a shaded area in Fig.~5 for {\it
Himiko} (the limits for {\it IOK--1} are very similar and are thus not
shown for clarity). With an upper limit to L$_{\rm FIR}$=
$\sim$4$\times10^{11}$\,L$_\odot$ our objects are the least
FIR--bright sources (by an order of magnitude) observed at high
redshift and start to probe the region that is only covered by local
galaxies at present.

We have shown above that the FIR--based SFRs can not be significantly
larger than the UV--based ones. If we assume that the FIR--based SFR
in these objects would equal the UV--based SFR , we can assign a
hypothetical FIR luminosity to both {\it Himiko} and {\it IOK--1}
(this luminosity would be lower by a factor of few if we had assumed a
dwarf SED template, see Fig.~3). This results in the (upper limit)
data points shown in Fig.~5. Under this assumption we would thus be
able to rule out high ratios such as L$_\cii$/L$_{\rm FIR}\sim$0.01
found in low-metallicity and low--luminosity nearby galaxies
(e.g. Israel et al.\ 1996, Madden et al.\ 1997). We stress however
that we have little constraints on L$_{\rm FIR}$ for both sources,
e.g. a very low L$_{\rm FIR}$ could result in significantly higher
L$_\cii$/L$_{\rm FIR}$ ratios.

\section{Concluding Remarks}

We have presented the results of a search for the \cii\ $^2$P$_{3/2}$
$\to$ $^2$P$_{1/2}$ line and the underlying continuum emission at
158\,$\mu$m in two LAE at high redshifts that form stars at moderate
rates (UV--based SFR $\sim$20\,M$_\odot$\,yr$^{-1}$), and a z$\sim$8.2
GRB. Our sources are the first to push studies of (unlensed) high--z
sources to FIR luminosities less than a few times 10$^{11}$\,L$_\odot$
and the resulting \cii/FIR limits begin to probe the luminosities and
parameter space occupied by local star--forming galaxies. The
continuum measurements in the LAEs rule out galaxy SEDs that show
extreme extinction, such as Arp\,220 and M\,82, and even those of
dust--rich nearby spiral galaxies such as M\,51. Based on our L$_{\rm
FIR}$ limits we derive that the (obscured) SFR cannot be higher than a
factor of $\sim$3 times based on the UV--derived SFR.  This correction
factor (to account for obscured star formation) is lower than the
typical correction factor of $\sim$5 to correct UV--based SFR for
extinction in lower--redshift LBGs (e.g., discussion in Pannella et
al.\ 2009). Our correction factor is also significantly lower than the
one derived for the Herschel PACS--detected subsample of LAEs at
2.0$<$z$<$3.5 where the total SFRs exceed the UV--based ones by more
than an order of magnitude (Oteo et al.\ 2012b, their  Tab.~1). However, our
finding is consistent with recent (rest--frame) UV studies of z$\sim$7
LBGs that also show only very little evidence for extinction in
systems of very similar redshifts (e.g. Bouwens et al. 2010).

Observations of the \cii\ line in galaxies at such high redshifts as
probed here will play a critical role in the future (e.g. Walter \&
Carilli 2008): the brightest line in the NIR (Ly--$\alpha$) is highly
resonant and will be affected by absorption in the intergalactic
medium (Gunn--Peterson effect) as well as possible dust aborption
within the galaxy. \cii\ emission, on the other hand, can in principle
provide key information on the star formation properties and dynamical
masses at these redshifts. The increase in sensitivity afforded by
ALMA will push the SFR limits obtained from \cii\ and FIR continuum
measurements down to only a few M$_\odot$\,yr$^{-1}$. We note however
that the lack of a sufficient number of ALMA `band 5' receivers
(163--211\,GHz) will not allow one to easily study galaxies emerging
from the Epoch of Reionization in the redshift range 8.0$<$z$<$10.65
with ALMA.

\acknowledgements

We thank the referee for critical and constructive comments that
helped to improve this paper.  We thank Dale Frail and Ruben
Salvaterra for useful discussions concerning GRB host galaxies. Based
on observations with the IRAM Plateau de Bure Interferometer.  IRAM is
supported by INSU/CNRS (France), MPG (Germany) and IGN (Spain). RD is
supported by DLR grant FKZ--50--OR--1104. DR acknowledges funding
through a Spitzer Space Telescope grant. We acknowledge the use of
GILDAS software (http://www.iram.fr/IRAMFR/GILDAS).

\end{document}